# Simulation-based Risk Reduction for Planning Inspections

Holger Neu[a], Thomas Hanne[b], Jürgen Münch[a], Stefan Nickel[b], Andreas Wirsen[b]

[a] Fraunhofer Institute for Experimental Software Engineering,
Sauerwiesen 6, 67661 Kaiserslautern, Germany
E-mail: `{neu, muench }@iese.fhg.de`

[b] Fraunhofer Institute for Industrial Mathematics,
Gottlieb-Daimler-Str. 49, 67663 Kaiserslautern, Germany
E-mail: `{hanne, nickel, wirsen}@itwm.fhg.de`

**Abstract.** Organizations that develop software have recognized that software process models are particularly useful for maintaining a high standard of quality. In the last decade, simulations of software processes were used in several settings and environments. This paper gives a short overview of the benefits of software process simulation and describes the development of a discrete-event model, a technique rarely used before in that field. The model introduced in this paper captures the behavior of a detailed code inspection process. It aims at reducing the risks inherent in implementing inspection processes and techniques in the overall development process. The determination of the underlying cause-effect relations using data mining techniques and empirical data is explained. Finally, the paper gives an outlook on our future work.

## 1   Introduction

The software industry is facing increasing demands to reduce time and costs for software products while increasing quality. Besides hiring better or more personnel and using more sophisticated tools and techniques, the improvement of the process itself is a crucial part of improving software development. Assessments and programs like CMM or SPICE address the maturity of the software development process. To achieve higher CMM or SPICE levels, the description and later the improvement of the process is important.

Usually, the key drivers for improving the software development process are time, cost, and product quality. Process improvement implies process change. Changing a software development process often leads to unpredictable results and therefore implies high risks. Occasionally, the change results in a worsening that is expensive to correct. Experienced personnel and a sufficiently valid software process model can reduce, but not eliminate, the risks. However, in practice, this only becomes evident after the implementation of the change.

Models help to understand the software development process and are a prerequisite for producing software with predictable quality and effort, but the models used and required often show only a static view of the process. The development of software is mainly a human-based activity and the related dynamic effects are often not considered when the process is changed.

Here, simulation can help to understand the current process and predict the effects of a process change before the change is implemented and the money is spent. Simulation can help to judge whether the change is an improvement of the process or, if several alternatives are available, to identify the alternative with the greatest benefit (e.g., selecting the appropriate inspection technique).

This paper is organized as follows: Section 2 gives an overview of simulation in the software engineering domain. In Section 3, we explain why we use a discrete-event model as modeling approach. Section 4 describes related work in this area. Section 5 explains our model and its goals. The input and the results of the simulation are discussed in Section 6. Finally, we give a summary and an outlook.

## 2   Background

In general, simulation can be used for planning a system a priori (e.g., before implementing it), for controlling a system (e.g., for operative or online usage), or for analyzing it (a posterior application). In the case of software development processes, a simulation model can support decision-making and risk reduction through a better understanding of the process. The reasons why a simulation model is created have been clustered in six categories of purpose [KMR99]: strategic management, planning, control and operational management, process improvement and technology adoption, understanding, training and learning.

The simulation model we have developed can be used for planning, process improvement, technology adoption and understanding. Planning has two perspectives; one perspective is thorough planning to make forecasts about such things as effort/cost, schedule and product quality, staffing levels needed, or resource allocation. The other planning perspective is to help select, customize and tailor processes for a specific context, e.g., for deciding which process alternative should be used, or which parts of the process can be skipped without quality loss. For the planning perspective, cost/effort, time and quality are often the dimensions of interest. However, for this purpose the model has to be more detailed than for other purposes, and it needs more precise input data.

Process improvement and technology adoption also have a planning perspective. Here, simulation can be useful if the process is to be changed according to suggestions, never tested in the considered environment, or technologies not used before. The decision is more difficult if several improvement alternatives are possible. Forecasting the impact with the simulation contributes to the improvement decision. If new technologies are to be introduced, at least some of the data has to be estimated or obtained from literature. The simulation results have to be compared with the actual process baseline (if available) or with alternative changes in order to decide on which change to implement.

In addition to making it easier to understand the software process, simulation helps to illustrate the process flow, the effects of feedback loops, and delays inherent in the process. Many project managers and experienced software professionals face problems when they try to explain these effects by themselves, which is due to their complicated dynamic behavior over time.

Before building a simulation model, the scope of the model has to be defined in accordance with the result variables, the process abstraction, and the input parameters. In general, the model scope (which usually fits one of the following definitions: portion of the life cycle, development project, multiple, concurrent projects, long-term product evolution or long-term organization) has two dimensions, time and organizational breadth. Time is the duration of the real process to be simulated and the breadth is the number of people or project teams involved [KMR99].

The result variables and the input parameters depend on the information and data available or needed. Typical result variables are: effort/cost, cycle time, defects, staffing, cost/benefit. Input parameters are the amount of work, defect detection efficiency, code rework effort, etc.

For the process abstraction, the model builder has to identify the key elements of the process, the relations between these elements and the behavior of these elements in the context of the simulation model. Obviously, the relevant elements are those necessary for fulfilling the purpose of the model. Important elements to identify are the key activities and tasks, objects (code units, designs, reports), resources (staff, hardware), dependencies of activities and flow of objects, loops and decisions (iteration, feedback), and other important interrelations.

To run a simulation model, the input parameters need to be initialized and the model has to be calibrated and validated to the target organization. Validation can be done through reviews and inspections of the model. However, in order to make a model fit an organization, the input data has to come from the organization. The quality of the simulation results depends on the accuracy of the input data. In an industrial setting the data is often not available because the metric data needed was not captured or different measures were collected. Useful strategies for handling these situations can be found in [KeR97]. Problems with the availability of data and also with the acceptance of simulation techniques when those are introduced are well known from other areas of application, but experience has shown that such difficulties can be overcome [McG98].

## 3    Approach

This paper focuses on the analytical processes in software development, especially the inspection processes. These processes are important for producing high quality products, and they are suited for building a simulation model, since a large amount of empirical data and knowledge exists for these processes. The simulation model can be used to play so-called "What If" games, for experimenting where real-life experiments are too expensive and time consuming, or even too dangerous or just impossible.

In the following, we explain why we use the discrete-event instead of the system dynamics approach for simulating the inspection-focused model.

Various simulation approaches have been used so far in the software engineering domain. These include system dynamics (or continuous simulation), discrete-event simulation, state-based process models, rule-based languages, and petri nets, among other approaches. System Dynamics (SD) is the simulation technique that was used in most cases for modeling software development processes. The first time SD was extensively used in that context was in the model of Abdel-Hamid and Madnick [AbM91]. After that, many SD models were built in the SE domain for different purposes. Regarding software inspections, a well-elaborated system dynamics model can be found in [Mad96] and [Mad94].

A few models [MaR00], [DoI01], [MaR01], [RCL99] use two simulation techniques, usually a continuous and a discrete approach, and are, therefore, called hybrid models. A pure discrete-event simulation was used in [ChS00].

[MaR00] give a good overview of the advantages and disadvantages of system dynamics and discrete-event simulation, especially if they are combined in a hybrid model.

System dynamics (SD) models usually focus on macro modeling levels and represent aggregated variables. The construction of an SD model of a software development process might be rather obvious by using nodes for representing aggregated variables such as total lines of code, total number of defects, or total effort. Arrows represent relationships between these variables. With SD models, it is easy to model feedback cycles caused by human effects such as fatigue, experience levels, schedule pressure or other dynamic influences. SD models are valuable for finding situations where models become unstable (because of feedback loops), and in predicting unanticipated side effects, if, for instance, the value of a variable is changed. However, explicit representations of objects such as team members or tasks to be done are not or not easily possible.

The nature of system dynamics models implies no specific process steps. If some design was done, coding starts immediately. To describe process steps such as "finish design before coding", additional mechanisms have to be modeled. In discrete-event simulation, a more detailed level with an explicit representation of objects is supported in a user-friendly way [BC84]. Thus, discrete-event models usually require higher complexity in modeling than SD models.

On the visual level, discrete-event models are more concrete than SD models because they represent static objects in reality by static blocks in the model, and transitory objects by moving units (MUs) similar to logistics simulation. In that area, MUs are, for instance, goods being transported by conveyors to a working station for processing them. SD models do not visualize such realistic operations but show logical relationships between the aggregated variables, in a way similar to that of cause-effect diagrams. While activities that occur in parallel are easier to model in SD models the idea of the simultaneous representation of one MU in two activities is difficult. For discrete-event models, the sequence of activities in a software process is easy to model, therefore this technique is extensively used for simulating manufacturing lines. The MUs flowing through the model represent items (design documents, code units) or persons. This representation is more flexible than representing MUs by static objects, especially with respect to a variable number of these objects. Discrete models allow individual attributes for each MU that can capture variations, for instance, in code size, difficulty, or personal productivity and experience. With these individual item attributes, we can determine, for example, an individual coding time for a code unit and a specific person. In addition, the values of the item's attributes can be sampled from distributions for capturing uncertainties in measuring the attributes.

The support of stochastic modeling is a seminal feature of tools for discrete-event simulation, e.g., by providing random number generators for various distribution functions. In SD tools, on the other hand, the representation of stochastic influences is usually less well supported.

In summary, SD is usually applied for rather general predictions, while discrete-event simulation is recognized as a reliable instrument for planning (e.g., in production and logistics). Since we focus on real-life applications of the model to specific industrial software development processes, we use the discrete-event simulation concept for our model. This provides us the possibility of characterizing the project items and persons by attributes. By doing so, the model can be used for a posteriori analysis with the data measured during the process. It is also possible to use historical data about items and persons to use the model to make predictions and planning for future projects.

The software package Extend has been chosen as a simulation tool for modeling software processes because it supports discrete-event modeling as well as time-continuous modeling in a systems dynamics fashion [Kra00]. Although we focus on building a time-discrete type of model analogous to logistics simulation, we also have the possibility to model specific parts of a software development process in a time-continuous way.

## 4   Related Work

Empirical studies about software inspections are an established discipline. A multitude of controlled experiments and case studies has been reported in literature (e.g., [LaD00]). Moreover, modeling and simulation are increasingly applied to software processes and widen their understanding. Raffo et al. [RKP+99] describe the multifaceted relationships between empirical studies and the building, deployment and usage of process and simulation models. Several models for simulating inspections are described. They mainly differ with regard to the intended purpose (e.g., prediction, control), the dependent variables of interest (e.g., cycle time, reliability), the development phases considered (e.g., design, all phases), the simulation technique, and the degree of combining simulation with other techniques that support process understanding (e.g., descriptive process modeling, GQM). In the following, some essential contributions are sketched.

Rus *et al*. [RCL99] present a process simulator for decision support that focuses on the impact of engineering practices on software reliability. The simulator consists of a system dynamics model and a discrete-event model. The continuous model is intended to support project planning and predict the impact of management and reliability engineering decisions. The discrete-event model is more suited for supporting project controlling. One main purpose of the discrete-event model is to predict, track, and control software defects and failure through out a specified period.

Madachy [Mad96] sketches a system dynamics simulation model of an inspection-based lifecycle process that demonstrates the effects of performing inspections or not performing them, the effectiveness of varied inspection policies, and the effects of other managerial decisions such as resource allocation. The model does not take into account schedule pressure effects and personnel mix.

Tvedt and Collofello [TvC95] describe a system dynamics model aiming at decision support with regard to several process improvement alternatives. The dependant variable of interest is cycle time. The model is intended for understanding cause-effect-relationships such as the influence of the implementation of inspections on cycle time reduction. The modeling approach distinguishes between a base model and several modular process improvement models (i.e., one for each improvement alternative).

Pfahl and Lebsanft [PfL99] combine process simulation techniques with static modeling methods, namely software process modeling and measurement-based quantitative modeling. They propose the IMMoS approach that integrates system dynamics modeling with descriptive process modeling and goal oriented measurement. The descriptive process model is used as a starting point for identifying causal relationships. Goal-oriented measurement is used for deriving measures from goals that are determined by the needs of a system dynamics model. Benefits of this combination are synergy effects from using already existing and proven methods and overcoming weaknesses of system dynamics model building.

In contrast to these contributions, the simulation model described in this paper focuses more strongly on the organizational and personal influence factors on inspections. It is mainly developed for risk reduction purposes. This model is much more detailed than other models described in the literature, and it is based on expert knowledge and partly on empirical data.

## 5   A Simulation Model for Analyzing the Effects of Inspections

A model is a more or less simplified representation of a real object, system, or any other subset of reality, but still similar with respect to certain aspects. It supports a better understanding and the analysis of the actual subject. Therefore, the model has to represent the essential information of the real process regarding static and dynamic relations. (See above) A guideline for developing discrete-event simulation models can be found in [Mus98].

In the following, we will show the development steps of the simulation model. At first we started with a static representation of the inspection model, which can usually be found in descriptive process models. These define the sequence of steps or activities, the products produced and changed, and the roles involved. For instance, an elicitation-based method [Bec01] and the tool SPEARMINT [SPE] can be used. The inspection process used for modeling is described in [EbS94].

To capture the dynamic behavior not described in a process model, we used, in a second step, cause effect (see Fig. 1) or casual loop diagrams [Ste00]. The variables of the cause effect diagrams depend on the goal of the simulation, e.g. ,the human effects like skills or productivities are important if effort, time and defects are of interest. The cause effect diagrams structure the assumptions of the dynamics inherent in the process.

In the next step, the static model was used to build the structure of the discrete event model while the cause effect diagram was used to define the elements of the equations for computing the variables of interest, e.g., coding time, defects, etc.

All variables appearing in the process model and cause effect diagrams of all sub-processes were classified with respect to their usage in an activity block or a moving unit. If a variable does not clearly belong to one of these groups it is classified as general process variable.

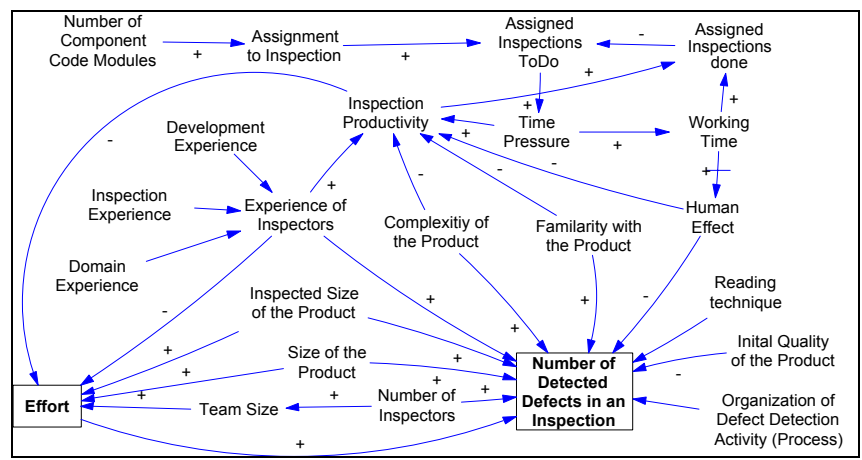

**Fig. 1.** Cause-effect diagram for effort and number of detected defects

The following groups were identified: Process variables, person variables, item variables, inspection specific variables, design, rework and coding specific variables. A person object, for example, consists of all attributes one would associate with humans, i.e., fatigue, skills in programming and inspection, personal productivity in coding and inspection, and a personal defect generation factor. General process variables include, e.g., information about factors for maximal productivity or quality measures used for computing the time an individual needs for performing an activity. The initial values of these attributes are usually inputs for the simulation model that have to be calibrated for a specific application of the model. A manager, for example, can use this freedom of choice of the parameters for studying the impact of introducing a software inspection into a concrete software development process in his or her company.

Now the functional dependencies between the different moving units in each sub-process have to be determined. It is obvious that these dependencies will differ with respect to the tasks of the MUs in the different sub-processes. For example, the interaction of persons and items (documents) depends on the process and on their role in it, i.e., in the coding sub-process a person is acting as a programmer, whereas in the inspection process, the same person can be an inspector or a moderator of the inspection of another item.

To specify the simulation model, the relationships qualitatively described in the cause-effect diagrams of all sub-processes and process models have to be quantified. Possible methods for quantifing the relationships are expert interviews, pragmatic models, stochastic analysis, and data mining techniques. Here one has to distinguish between the quantification of known relationships, i.e., exact linguistic descriptions that are available and have to be transformed to mathematical functions, and the generation of new rules by applying data mining techniques to describe relations that are not obvious or were not considered up to now in simulations of the software development process. The choice of the data mining techniques for quantification depends on the data or information available from experiments. The data will also be used for the validation of functional relations known from literature.

Up to now we have been using neural networks and classification trees for the quantification of the qualitatively known relations. The neural networks used are feed forward neural networks with one hidden layer and a single output [Whi89] [Whi90]. The neural network function is given by

$$f(x,\Theta) = v_0 + \sum_{k=1}^{n} v_k g(x w_h) \tag{1}$$

where $\Theta=(w_1',\ldots w_h',v_0,\ldots v_h)$ is the vector of network weights, while $w_h'=(w_{0h},\ldots,w_{dh})$, $h=1,..,n$, g is the activation function from each hidden unit and $x=(1,x^1,\ldots x^d)$ is the vector of inputs. The network weights in our simulation model were estimated from industrial data.

Additionally, relevance measures [Sar02] were used to determine the impact of each explaining (input) variable with respect to the explained (output) variable. The relevance measure here, denoted as $R(f,x)$, was computed as the partial derivatives of the trained network function $f(x,\Theta)$ with respect to explaining input variables $x^k$, i.e.,

$$R(f,x^k) = \frac{\partial f(x,\Theta)}{\partial x^k}. \tag{2}$$

Applying these measures, one can easily verify whether the estimated functional dependencies describe the input/output relation in a sufficient manner, or whether an explaining variable is missing. In the latter case, a new rule has to be generated for applying data mining techniques. Thus, a relevance measure can also be used to validate the qualitative description of the dependencies given in the cause-effect diagrams.

The second technique applied for the quantification of the relationships we used are classification trees determined by the software tool XpertRule Miner [Xpe]. Based on the information gain technique, a classification tree for an explained variable is calculated on the data set. The tree can be read from root to a leaf as a rule for the input/output relation. A leaf of the tree contains information about the percentage, variance, mean and standard deviation for the explained variable when applying the corresponding rule. Thus, the root of the tree denotes the variable with the greatest impact with respect to the explained variable. The splitting criterion used is based on the normalized standard deviation [Xpe].

In the following, we will describe the simulation model, which is being developed for decision support on organizing inspection processes. We focus on the variables size and percentage of modules to be inspected, number of inspectors and inspection effort.

The sub-processes of the process diagram pre-structure the sequence of tasks (design, coding, inspection, test, etc.) to be performed during a software development process. Similar to lines in material flow systems the activity blocks and other modeling blocks are connected for guiding items and persons through the system.

The general idea of directing MUs through the system is as follows: Before specific tasks can be performed, items and persons have to be assigned. In Extend, a block creating a compound MU, which represents an item together with an assigned person, supports this. Up to now, one main assumption of the model is that the assignment of tasks to persons is done in an arbitrary way or, more precisely, persons are selected from a staff pool in a "first come, first serve" fashion.

After accomplishing a task, linked items and persons are un-batched. Persons are then directed back to the pool where they are waiting for new tasks to be assigned to them. Items are directed to the subsequent sub-process, e.g., from design to coding or from inspection to rework. In some cases, there are alternatives for routing an item. For instance, rules may be applied for deciding on whether an item is subject to inspection or not. Moreover, switches can be used for activating or deactivating some parts of the model, e.g., the inspections, the design, or the testing and rework activity. In general, the connections of processes and sub-processes and the routing logic for the MUs represent organizational rules of a considered real-life company. Because of the complex nature of the processes to be modeled, it is hardly possible to represent many details of a real-life software development process. For instance, meetings prior to an inspection are not represented in the model. Let us note that the sub-processes outside our model focus are modeled in a rougher way. Similarly, aspects outside the considered project are neglected, for instance the involvement of persons in other tasks. It would, in principle, be possible to have a pre-specified timetable for each person defining the availability of the person for the current project.

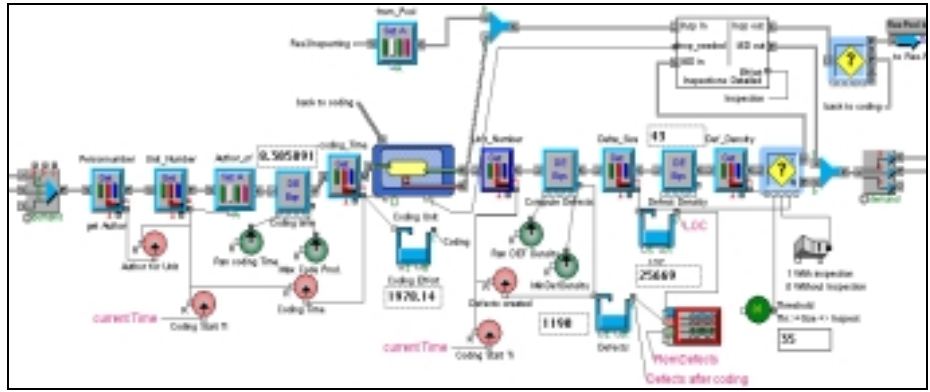

**Fig. 2.** Excerpt from the Extend visual Interface of the model. Coding with the inspection of selected code items

Most of the blocks used for representing a sub-process (see Fig. 2) are for accessing attributes or variables and for calculating and assigning new values. For instance, in the coding process, the number of defects produced is calculated; in the inspection and test processes, the number of defects found is calculated, and in the rework processes, the number of defects is updated (considering new defects produced during rework). The most important block of a sub-process is, however, a working unit representing the activity of that sub-process in a narrower sense. When a compound MU reaches this block, it stays there for a specific time. This working time has to be calculated beforehand.

The most important issue in each sub-process is to represent the quantitative relationships of the model in a valid way, especially those output values that are most interesting for the user, i.e., changes in the number of defects and time consumed by the various activities. In general, these values are determined by attributes of the items, by attributes of the persons, and by general or project-specific factors. For some of the relevant data it is hardly possible to determine the necessary information in real-life projects. For instance, details on the specific experiences, skills, and productivities of persons are usually not available. Therefore, we have elaborated approaches for taking such human attributes into account, which are not directly observable, and for considering them in the quantitative logic of the model. For instance, a logistic learning model has been adapted and implemented in the simulation model. This is based on various skills that increase with experience and that affect the (ex post) productivity of a person. In a similar way, a model for considering time pressure and its influence on the productivity has been worked out.

Even after careful refinement of the quantitative relations, there are essential effects in a software development process that cannot be fully determined a priori, for instance, human effects such as fatigue, boredom, and other physical and mental factors. These human effects are be considered by having stochastic elements in the process, which influence, for instance, the working times and numbers of defects produced.

As input data, the simulation model requires a specification of a software development project. Roughly said, such a project consists of item-specific data, person-specific data, and project-specific or general data. For each item to be produced (e.g. the source code of a module), the item-specific data includes the number of lines of code (or new/changed lines of code in case the item existed before) and a complexity measure assumed to be around 1. The person-specific data consists of estimates for skills or productivities of all members of the development team.

General or project-specific data are, for instance, average and maximum productivity values or person costs per hour. Alternatively, it is possible to generate a stack of tasks (items) and a pool of persons randomly for applying the model unrelated to a given project, e.g., for educational purposes. Input and output data are stored in a text file linked via SDI Interface with the Extend simulation software. An internal DB stores values for used distributions that can be changed if the model is to be fitted for a specific context.

When the input data are defined, the simulation can simply be started and runs automatically. If the animation is switched on, one can observe people and items moving through the various sub-processes represented in the model similar to a logistics simulation. The overall information value of the animation is, however, rather small. Worthwhile tools for keeping the user informed about the dynamics of the model are plotter blocks, which show the charts of specified variables. Additionally, various variable values are displayed within the window of the simulation model. During the model run, values of various variables are written to the file for further processing. These output variables are, for instance, the number of defects of each item produced during coding, found during inspections, produced during rework and so on, data on the required working times, and effort spent for each task.

If the user is mainly interested in the results of a simulation run, he or she may just consider the output data. In that case, it is possible to switch off the time-consuming animation for getting the results more quickly. It is then possible to use such a model in an online fashion, e.g., for quickly analyzing the effects of varying model parameters. It is planned to equip the model with a simulation cockpit for providing a user-friendly, custom-made interface to the simulation model.

## 6    Results of the Simulation Model

In this Section, we use the simulation model for two experiment series on variants of the software development process. In both series, the objectives 'duration' and 'overall effort' are considered. For facilitating comparisons, the third objective, product quality, is assumed to be constant. This is achieved by requiring the test phase to continue until a minimum level for the defect density is reached. This means that products with more defects entering the test phase require more test and rework effort.

For the sub-processes testing and rework, it is calculated how much time the test activity requires to get a desired defects density. Thus, the rework effort depends on the number of defects to be found in test, and after testing, the resulting number of remaining defects is always the same.

In 6.1, we analyze whether inspections of all or selected items are useful compared to a software development process without inspections. In 6.2, we analyze the question of what an optimal size of an inspection team might be.

Of course, there are further parameters of the simulation model that influence the effectiveness of inspections and that could be analyzed by the simulation model, e.g., the inspection technique. Corresponding studies based on the simulation model will be performed in the future.

### 6.1 The Selection of Items for Inspection

One of the key questions in introducing and planning inspections concerns the selection of items to be inspected. In order to compare different policies, we look at a project for producing software (creating new features for an existing product) with 100 items of different size, with 20 developers, and compare the overall effort and time spent for a specific defect density. We consider three variants of a software development process: a) without any inspections, b) with inspecting all items, and c) with inspecting all items with a defect density larger than 35 defects per KLOC. This defect density threshold turned out to be reasonable according to the given defect distribution. Usually the number of defects in a piece of code is not known. This assumption can give a baseline for the effects of an optimal selection of code units for the inspection. Alternatively, other rules could be used, for instance, rules based on the size of the document.

Table 1 shows the simulation results for 100 items with a purposed defect density of 1.5 defects per 1000 lines of code. The model shows that the introduction of inspections increases the effort spent for the coding phase but if the inspections are executed, the effort spent for testing and rework is reduced.

| Alternatives | Size | Defects coded | Defects found during inspection | Defects after insp. | Defects | Overall Effort | Duration |
|---|---|---|---|---|---|---|---|
| No inspections | 25669 | 1135 | 552 | 485 | 33 | 9885.64 | 732.7981 |
| All inspected | 25669 | 1149 | 650 | 564 | 33 | 9538.59 | 614.6752 |
| Select item for inspection | 25669 | 1137 | 578 | 503 | 33 | 9534.09 | 619.104 |

**Table 1.** Average results of 20 simulation runs

The overall effort is less for the simulation runs with inspections. Also, the duration of the project is shorter if inspections are executed.

These results are in accordance with the literature [FrW90], which suggests that the effort spent for inspections is less than the effort saved for testing. The results of the simulation runs for items selected for inspections show that with such a policy the overall effort can be further decreased, but only with some increase the project duration.

As an alternative to the used testing policy, it would be possible to specify a time frame for how long a code item is tested (instead of specifying a desired defect density) so that the effort spent on testing could be kept constant. In that case, software development processes with inspections would result in better product quality at the end of the software development process.

### 6.2   The Influence of the Team Size

The number of found defects and the effort in the overall process (especially coding and test) depend on the number of inspectors involved in the inspection of one code item. For analyzing these effects we perform simulation runs of the model with the size of the inspection team varying from 1 to 10.

In Fig. 3 the graph shows the overall effort needed for the process. It significantly increases for more than four inspectors. Similar increases are shown for the duration for the overall process. If we consider the effort and the duration, the optimal number of inspectors is between two and four.

The other two lines in the graph show the number of defects found and missed during the inspection. Here we can see that increasing the number of inspectors does increase the number of defects found, but only degressively. With more than seven inspectors the number of found defects does not increase significantly. Therefore, an inspection team size of more than seven inspectors is not useful for increasing the product quality.

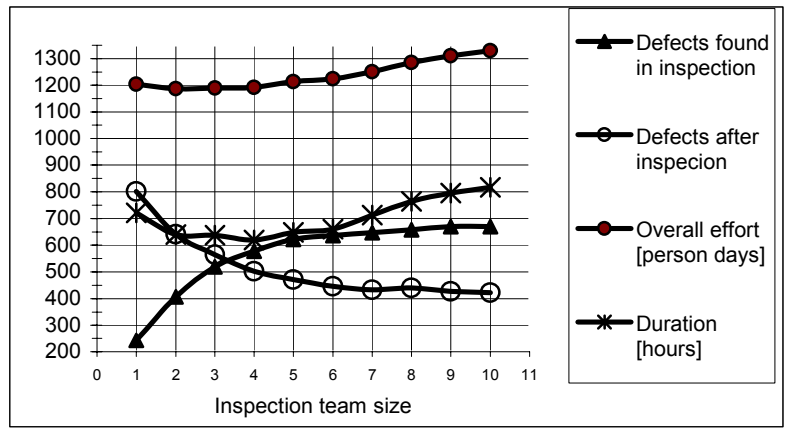

**Fig. 3.** Defects, duration and effort with respect to different numbers of inspectors (average values for 20 simulation runs).

As stated in [LLS+99], increasing the number of reviewers has a ceiling effect because the probability that defects are found by two or more inspectors increases with the number of inspectors. Therefore, adding inspectors does not increase the number of defects detected significantly and mainly increases the effort and time spent.

## 7   Summary and Outlook

In this article, we have described the development and usage of a simulation model for software development processes. This model can be applied for various purposes such as improving the understanding of these processes, and for planning in an industrial setting. Unlike most existing work in that area, our model is based on discrete-event simulation and supports a more detailed representation of analytical processes with a focus on inspections. Techniques used for developing the model were static process models, cause-effect diagrams, general models documented in the literature, and tools for data analysis. As we have seen, the questions of calibrating and validating are among the main issues for modeling.

For this purpose, we have used various kinds of data such as data from industrial companies, but also general results taken from the relevant literature. In practical settings, data required for model calibration is usually not complete, in particular with respect to human effects, i.e., experience and fatigue. Because of this reason, the modeling of stochastic effects is a significant feature of the model, especially for adequately considering the risks when changing the software development processes in a company. In any case, the employment of rules generated by advanced data mining tools and techniques like fuzzy logic and rough sets for rule generation will be an important issue for further development of the model.

Some exemplary applications of the model led to results similar to those expected from the literature, i.e., emphasizing the usefulness of inspections in software development. We cannot directly compare the overall effort and duration with the defects found during inspection, because the test policy (see above) is not elaborated enough.

For future development of the model, we plan to apply some optimization tools especially for calculating a load balancing assignment and for scheduling staff in an online fashion [FW98, LN02], which should lead to more efficient results than the current arbitrary mechanism used in the simulation model.

It is planned to apply the model in an industrial application for a case study. For this purpose, further refinements of the model will include the provision of modular models and the modeling of other sub-processes that are currently not or just roughly included in the model.

## Acknowledgements

The main work for this paper was done in connection with the SEV project and the ProSim project. The projects where supported by the German Bundesministerium für Bildung und Forschung (SEV) and the Stiftung Rheinland-Pfalz für Innovation (ProSim, project no.: 559)

We would like to thank Sonnhild Namingha from the Fraunhofer Institute for Experimental Software Engineering (IESE) for reviewing the first version of the article.

## References

[AbM91]   T. Abdel-Hamid, S. E. Madnick: Software Project Dynamics. An Integrated Approach. Prentice Hall, Englewood Cliffs 1991.

[Bec01]   U. Becker-Kornstaedt: Towards systematic knowledge elicitation for descriptive software process modeling. F. Bomarius, S. Komi-Sirviö (Eds.): Proceedings of the Third International Conference on Product–Focused Software Processes Improvement (PROFES), Kaiserslautern, September 2001. Lecture Notes in Computer Science 2188, Springer, Berlin 2001, 312–325.

[BC84]    J. Banks, J. S. Carson, II: Discrete-Event System Simulation. Prentice-Hall, Englewood Cliffs 1984.

[ChS00]   A. M. Christie, M. J. Staley: Organizational and social simulation of a software requirements development process. Software Process Improvement and Practice, 2000, 103-110.

[DoI01]   P. Donzelli, G. Iazeolla: Hybrid simulation modelling of the software process. Journal of Systems and Software 59, 3, 2001, 227-235.

[EbS94]   Ebenau, Robert G.; Strauss, Susan H.: Software Inspection Process. New York: McGraw-Hill, Inc., 1994.

[FG98]    A. Fiat, G. J. Woeginger (Eds.): Online Algorithms: The State of the Art, Springer, Berlin 1998.

[KeR97]   M. Kellner, D. Raffo: Measurement issues in quantitative simulations of process models. Proceedings of the Workshop on Process Modelling and Empirical Studies of Software Evolution (in conjunction with the 19th International Conference on Software Engineering), Boston, Massachusetts, May 18, 1997. 33-37.

[KMR99]   M. I. Kellner, R. J. Madachy, D. M. Raffo: Software process simulation modeling: Why? What? How? Journal of Systems and Software 46, 2-3, 1999, 91-105.

[Kra00]   D. Krahl: The Extend simulation environment. J.A. Joines, R. R. Barton, K. Kang, P. A. Fishwick (Eds.): Proceedings of the 2000 Winter Simulation Conference. IEEE Press, 2000, 280-289.

[LaD00]   O. Laitenberger, J.-M. DeBaud: An encompassing life-cycle centric survey of software inspection. Journal of Systems and Software 50, 1, 2000, 5-31.

[LEH99]   O. Laitenberger, K. El Emam, T. Harbich: An Internally Replicated Quasi-Experimental Comparison of Checklist and Perspective-Based Reading of Code Documents. IEEE Transactions on Software Engineering 27, 5, 2001, 387-421.

[LeR99]   M. M. Lehman, J. F. Ramil: The impact of feedback in the global software process. Journal of Systems and Software 46, 2-3, 1999, 123-134.

[LN02]    A. Lavrov, S. Nickel: Simulation und Optimierung zur Planung und Steuerung von Kommissioniersystemen. VDI-Wissensforum Optimierte Kommissioniersysteme, March 2002, K. 10, 1- 16.

[Mad94]   R. J. Madachy: A Software Process Dynamics Model for Process Cost, Schedule and Risk Assessment, PhD Dissertation, Department of Industrial and Systems Engineering, USC, December, 1994.

[Mad96]   R. J. Madachy: System dynamics modeling of an inspection-based process. Proceedings of the Eighteenth International Conference on Software Engineering, IEEE Computer Society Press, Berlin, Germany, March 1996, 376-386.

[MaR00]   R. H. Martin, D. Raffo: A model of the software development process using both continuous and discrete models. Software Process Improvement and Practice, 2000, 147-157.

[MaR01]   R. Martin, D. Raffo: Application of a hybrid process simulation model to a software development project. Journal of Systems and Software 59, 3, 2001, 237-246.

[McG98]   F. McGuire: Simulation in healthcare. J. Banks (Ed.): Handbook of Simulation. Wiley, New York 1998, 605-627.

[Mus98]   K. J. Musselman: Guidelines for success. J. Banks (Ed.): Handbook of Simulation. Wiley, New York 1998, 721-743.
[PfL99]   D. Pfahl, K. Lebsanft: Integration of system dynamics modelling with descriptive process modelling and goal-oriented measurement. The Journal of Systems and Software 46, 1999, 135-150.
[RCL99]   I. Rus, J. Collofello, P. Lakey: Software process simulation for reliability management. Journal of Systems and Software 46, 2-3, 1999, 173-182.
[RKP+99]  D. Raffo, T. Kaltio, D. Partridge, K. Phalp, J. F. Ramil: Empirical studies applied to software process models. International Journal on Empirical Software Engineering 4, 4, 1999, 351-367.
[Sar02]   A. Sarishvili: Neural Network Based Lag Selection for Multivariate Time Series. Phd. Thesis, University of Kaiserslautern, 2002.
[Ste00]   J. D. Sterman: Busines Dynamics – Systems Thinking and Modeling for a Complex World, Irwin McGraw-Hill, 2000.
[SPE]     The SPEARMINT web site: www.iese.fraunhofer.de/Spearmint_EPG.
[TvC95]   J. D. Tvedt, J. S. Collofello: Evaluating the effectiveness of process improvements on software development cycle time via system dynamics modeling. Proceedings of the Computer Software and Applications Conference (CompSAC'95), 1995, 318-325.
[Whi89]   H. White: Learning in artificial neural networks: A statistical perspective. Neural Computation 1, 1989, 425-464.
[Whi90]   H. White: Connectionist nonparametric regression: multi layer feed forward networks can learn arbitrary mappings. Neural Networks 3, 1990, 535-549.
[Xpe]     XpertRule Miner, Attar Software GmbH: www.attar.com.